\documentclass[12pt]{article}

\usepackage[utf8]{inputenc}
\usepackage[T1]{fontenc}

\usepackage[a4paper, margin=2.7cm]{geometry}
\usepackage{graphicx}
\usepackage{subcaption}
\usepackage{amsmath, amsthm, amsfonts, amssymb}
\usepackage{mathrsfs}           
\usepackage[colorlinks=true,linkcolor=blue,citecolor=blue,urlcolor=blue,breaklinks]{hyperref}

\usepackage{algorithm}
\usepackage{algpseudocode}

\usepackage{bbm}

\title{Degree-Based Random Walk Approach for Graph Embedding}

\author{  Sarmad N. MOHAMMED \\ 
           \small Computer Science Department, \\
           \small  Kirkuk University,  \\
           \small  Kirkuk, Iraq \\
           \texttt{\small sarmad\_mohammed@uokirkuk.edu.iq}
           \and 
           Semra G\"UND\"U\c{C} \\ 
           \small Computer Engineering Department, \\
           \small  Ankara University,  \\
           \small  Ankara, Turkey \\
           \texttt{\small Semra.Gunduc@ankara.edu.tr}
        }

\date{}

\begin{document}

\maketitle

\begin{abstract}

Graph embedding, representing local and global neighbourhood information by numerical vectors, is a crucial part of the mathematical modeling of a wide range of real-world systems. Among the embedding algorithms, random walk-based algorithms have proven to be very successful. These algorithms collect information by creating numerous random walks with a predefined number of steps. Creating random walks is the most demanding part of the embedding process. The computation demand increases with the size of the network. Moreover, for real-world networks, considering all nodes on the same footing, the abundance of low-degree nodes creates an imbalanced data problem. In this work, a computationally less intensive and node connectivity aware uniform sampling method is proposed. In the proposed method, the number of random walks is created proportionally with the degree of the node. The advantages of the proposed algorithm become more enhanced when the algorithm is applied to large graphs. A comparative study using two networks, namely CORA and CiteSeer, is presented. Compared with the fixed number of walks case, the proposed method requires approximately $50 \%$ less computational effort to reach the same accuracy for node classification and link prediction calculations.\\

  
\end{abstract}

\section{Introduction}
\label{sec:intro}

Networks are ubiquitous and are the main infrastructure for modeling biological, physical as well as social systems. In any particular event, the determination of the roles of the nodes becomes crucial for both better understanding the dynamics and, if necessary, taking the prevention measures (for example in the epidemic case who spreads the disease faster, in the social network who influences the others, etc.). Hence the networks are crucial tools for modeling and understanding the real-world systems by the introduction of a device to map elements of the graph into a low-dimensional representation. To achieve this goal, it is important to preserve the local and global properties of individual nodes in a very wide range of graph structures that are seen in real-world and artificial networks.

$\;$

Mathematical modeling of such a wide range of phenomena requires a good description of the underlying connectivity structure among the nodes. The connectivity of the nodes determines the pattern of the interactions and dynamics. For example in social networks, to recommend a new friend or to predict the role of a person needs different approaches mainly sourced from the topology, like considering the number of common neighbors, strength of the link, or being in the same community.  Main tasks include node classification \cite{bhagat2011node, kumar2020node}, link prediction \cite{liben2007link, martinez2016survey}, anomaly detection \cite{chandola2009anomaly, ma2021comprehensive}, and community detection \cite{fortunato2010community}. 
Many machine learning algorithms \cite{witten2005practical, kotsiantis2007supervised, jordan2015machine} have shown to be very successful in prediction, classification, reconstruction, and various more complicated tasks. For the success of each task specially prepared input data is necessary. Graphs are different data structures. They have richer features than usual inputs of machine learning algorithms such as pictures or categorical items. Hence, an algorithm to capture the information behind this high dimensional data to introduce the machine learning algorithms as the input vector is necessary.

$\;$

Traditional approaches for extracting the structural features of graphs use statistical techniques \cite{huang2010link, fortunato2004method, gallagher2008leveraging, henderson2011s, murata2008link} (number of neighbors, clustering coefficient, centrality measures, etc.). Traditional methods are time-consuming and they are short of fully exploiting the hidden features of graphs. Recently a new idea has been employed to embed the nodes into a low dimensional vector space by using machine learning techniques, namely representation learning. In the literature, a wide range of approaches has been presented (\cite{makarov2021survey} for a survey). The method imported from natural-language processing has been employed in embedding the connectivity structure of graphs into n-dimensional vectors. This approach uses the similarities between the structure of natural languages and the connectivity structure of graphs. 

$\;$

Some of the successful methods use random walks to collect the nodes' local and global connectivity structures. These models have attracted much attention \cite{perozzi2014deepwalk, grover2016node2vec, tang2015line, ribeiro2017struc2vec, nguyen2018biasedwalk}. The idea behind the use of random walk statistics is to enlist the co-occurrence of nodes. The neighbors of a given node are visited on the repeated random walks. The co-occurrence of the nodes in the same random walk ensures the connectivity of the group of nodes. The information collected during random walks is used in analogy with the natural language processing algorithms. Well-known random walk-based embedding methods are DeepWalk \cite{perozzi2014deepwalk} and node2vec \cite{grover2016node2vec}. Both methods are shallow embedding and follow an optimization strategy by using co-occurrence statistics.

$\;$

Many recent studies are using the random walk method. These studies perform various network analysis tasks by reducing the number of training samples and computational cost with various random walk approaches. 
In \cite{ahmed2019role2vec, ma2019riwalk}, nodes are embedded based on specified characteristics of random walks that begin at one node and proceed to the next. Role2vec \cite{ahmed2019role2vec} samples a corpus using attributed random walks, which preserves each node's structural type as well as increases the efficiency of available space. Based on the structural patterns of nodes in each surrounding, RiWalk \cite{ma2019riwalk} learns an embedding for each node by building a relabeled subgraph for each node and performs random walks on the subgraph created.
As a scalable graph embedding approach, DiaRW \cite{ zhang2019degree} has recently been proposed as one that uses a degree-biased random walk and variable lengths policy. DiaRW method creates random walks depending on the source node's centrality in reducing sampling redundancy. 
To modify the node preferences when walking, BalNode2Vec \cite{salamat2020balnode2vec} defined the concept of a node's network neighborhood and developed a balanced random walk process that adapts to the graph's structure.
Fazaeli and Momtazi \cite{fazaeli2022guidedwalk} presented GuidedWalk, the idea behind this method is to increase the probability of visiting the local structure of nodes in the same class by using the label information of the nodes in the random walk phase. Wang et al. \cite{wang2022hashwalk} proposed the HashWalk method to preserve the network topology and improve node embedding. This method first compresses the cliques in the network into single nodes and then obtains the compressed clique sequences using the random walk method.
However, even though the state-of-the-art random walk-based embedding methods have accomplished some benefits, they cannot disregard the restrictions imposed by randomness in the walking process. For example, eliminating oversampling (number of random walks being chosen proportional to the degree of the node).

$\;$

The embedding strategy guarantee that the similarities of the nodes are related to the similarities of corresponding embedding vectors. Both node2vec and DeepWalk algorithms use the same random walk approach to collect information from the graph. The difference between DeepWalk and node2vec is that they use different optimizations and approximations to compute the embeddings. Besides that, the two methods use the same user-defined initial values for the number of walks, and the walk length to capture the features. These values are set by the user at the beginning of the data collection stage and optimized values can only be achieved with trial and error.

$\;$

Among the real-world networks, scale-free networks have a special place since they represent the majority class among the real-world networks \cite{barabasi2003scale}. The characteristic power-law degree distribution of the scale-free networks is an indication of many nodes with a low degree and a few nodes with a high degree. For this reason, the fixed number of random walks oversample the low degree while high degree nodes can not be sampled as necessary. In case of all nodes are considered on equal footing, to increase the sampling rate of high degree nodes, it is necessary to increase the number of walks that start from every node. This situation corresponds to an increase in the computation time proportional to the size of the network. Meantime it results in an excessive number of walks for low degree nodes. This situation results in the creation of imbalanced data during the embedding stage.

$\;$

The present work aims to propose a modification in the feature extracting algorithm.  To eliminate oversampling, the number of random walks is chosen proportional to the degree of the node. Low degree nodes are sampled relatively less than higher degree nodes. Considering the proportions of the low and high degree nodes in scale-free networks,  this choice reduces the total number of walks considerably while increasing the sampling rate of the high degree nodes. This modification introduces remedies for excessive computational time with the growing network size and it also helps to overcome imbalanced data problems. In the proposed model the number of walks that start from each node, tuned to be proportional with the degree of the related node instead of a fixed number, while for the rest of embedding the node2vec strategy is employed.
The paper is organized as follows: the next chapter is devoted to the explanation of the method. The results of the comparative study of the proposed algorithm are presented in the third section. The final section consists of the discussions over the results and conclusions.

\section{Method}
\label{Method}

The successful implementation of representation learning in natural language processing domains (e.g., word2vec \cite{mikolov2013distributed, mikolov2013efficient}) paved the way for new directions in network representation learning by optimizing the neighborhood preserving likelihood concept. A document is a collection of words. In word2vec the features of the words are extracted from the relative occurrence with the related, or in other words, nearby words (being in the same window size). A text document consists of already existing sentences. The sentences are the natural constructs representing the relations among the words.  There are no such sequences or grouping for the nodes of the networks. The difficulty of using representation learning techniques on the graphs appears to be constructing sentence-like structures. To obtain such structured constructions which introduce an order for nodes and reveals hidden features, random walks are the best candidates and often employed.

$\;$

\begin{figure}[h!]
\begin{center}
    \includegraphics[width=9.0cm]{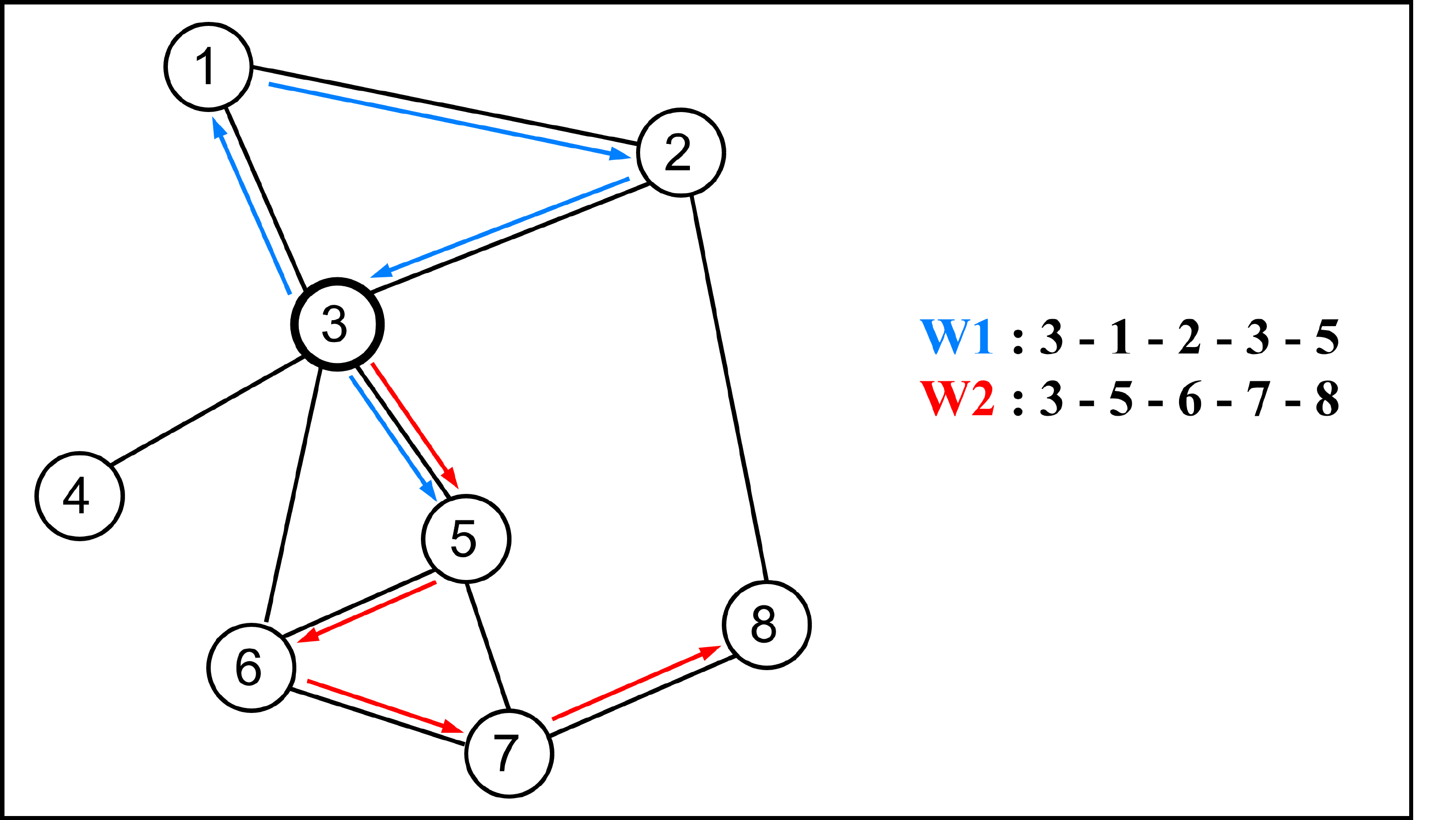}
    \caption{A representation of 2 random walks starting from node  $n_3$  on a sample graph.}
    \label{fig:1}
\end{center}
\end{figure}

Graphs, ($G$) essentially consist of two components; nodes ($N$) and edges ($E$). Nodes are entities that are connected by edges. There is a neighbor set for each node ($n_i$) in the graph. These neighbors will be determined by a set of random walks with some strategy. This strategy must guarantee that local and global neighborhood information is integrated into the embedding. As shown in Figure \ref{fig:1}, third node  has $5$ neighbors, $N_3  = \{n_1,n_2 ,n_4 ,n_5, n_6 \}$, and if we start two random walks from the node $n_3$,  following the same strategy (the walk length is  $4$), then the first walk ($W_1$) can be ($n_3 \rightarrow n_1 \rightarrow n_2 \rightarrow n_3 \rightarrow n_5$ ) and the second walk ($W_2$) can be ($n_3 \rightarrow n_5 \rightarrow n_6 \rightarrow n_7 \rightarrow n_8$ ). These walks will give sentence-like sequences:

\begin{eqnarray}
\;\;&S_1:& n_3 \; n_1\; n_2\;  n_3\; n_5. \nonumber \\
\;\;&S_2:& n_3 \; n_5 \; n_6 \; n_7 \; n_8.\nonumber
\end{eqnarray}

The idea behind creating sentence-like node sequences is to represent the similarities between the nodes with the corresponding embedding vectors. Every random walk produces a sequence that possesses the inherent hidden traits between the nodes. The walk length enables to penetrate deeper into the graph while the number of walks adds different connectivity patterns. Hence, more accurate embedding vectors can be obtained.  So there must be enough walks with sufficient length to map the different paths start from each node.

$\;$

To detect the similarity of the nodes in a graph both connectivity and structural resembles must be taken into account. In the node2vec model, the parameters $p$ and $q$ allow sampling local and global neighborhoods of nodes in the walks. The parameter $p$ controls the probability of returning to the previous node while  $q$ value makes visiting the far-away neighbors possible. For low $p$ values, it results in revisiting already visited neighbors for through the graph.

\begin{figure}[h]
\begin{center}
    \includegraphics[width=15.0cm]{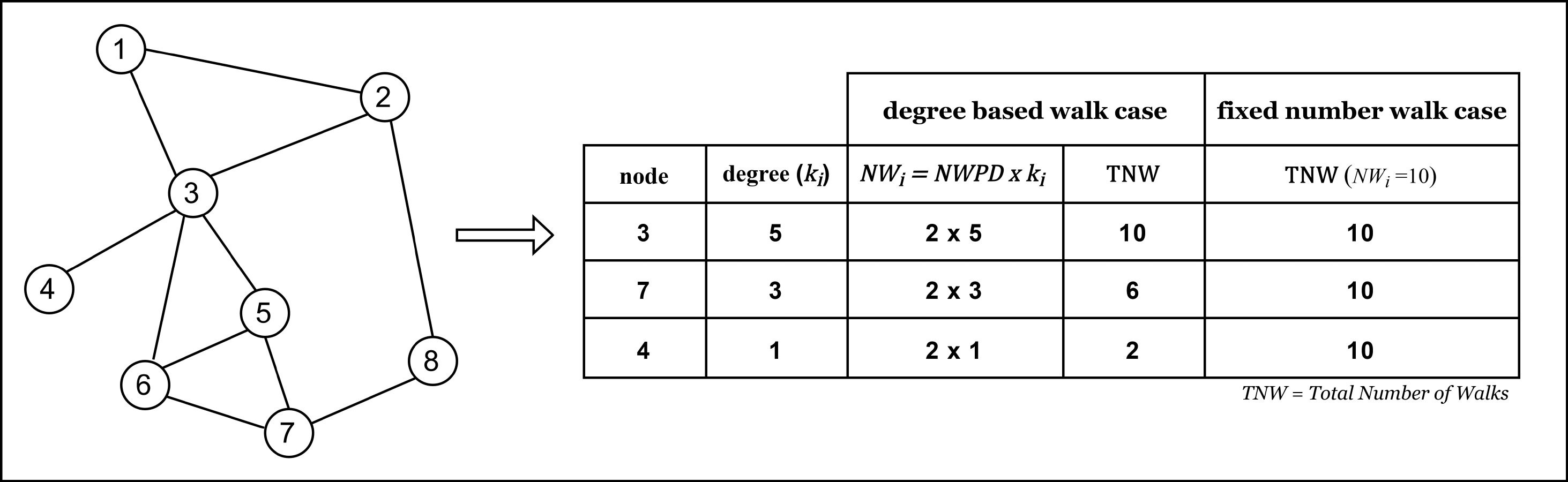}
    \caption{An illustration of counting differences between degree-based and fixed number of random walks.}
    \label{fig:2}
\end{center}
\end{figure}

In a fixed number of walk cases, the same number of walks start from each node, respectively. In the case where the number of walks proportional with the degree of the node, the number of walks started from a node is given as,
\begin{equation}
{ N\!W}_i = N\!W\!P\!D \times k_i,
\end{equation}
where ${ N\!W}_i$ and $k_i$ are the number of walks started  and the degree of the $i^{th}$ node.  ${ N\!W\!P\!D}$ is  the number of walks per degree. Figure \ref{fig:2} presents an illustration of the difference in random walk strategy between degree-based node2vec and node2vec in a sample network.

\begin{algorithm}
\caption{Degree-Based Sampling}\label{Pseudocode}
\begin{algorithmic}

\State \textbf{Input:}  Network $G = (V,E)$
\State  \textbf{Initialization:}\\
\quad Number of nodes $N$;\\
\quad Array of node degrees $D$;\\
\quad Length of the walk $WL$;\\
\quad Number of walks per degree $WD$;\\
\quad Return-weight parameter $p$;\\
\quad Neighbor weight parameter $q$;\\

\State \textit{\textbf{*Section to create degree based random walks per node:}}

\textbf{for} node in $N$ \textbf{do}        \Comment{Loop over the nodes}\\

\quad \quad \quad \quad StartNode = node    \Comment{Starting node}\\
\quad \quad \quad \quad epochs = $WD$ * $D$[node]   \Comment{How many times to start a walk from the node}\\

\quad \quad \quad \quad Walk[node] = Random Walk($G$, StartNode, $WL$, epochs, $p$, $q$)\\
			
\quad \quad \textbf{endfor} 

\end{algorithmic}
\end{algorithm}

Algorithm~\ref{Pseudocode} shows the pseudocode for creating random walks per node using degree-based weighted sampling approach. This approach provides more even sampling  the network,  since the nodes have varying number of neighbors. Moreover, for large networks, it increases the sampling efficiency. For large scale-free networks, the probability distribution is given by,
\begin{equation}
  P(k) = (\gamma - 1)\times k_{min}^{\gamma -1} \times k^{-\gamma}
  \end{equation}
where $k_{min}$ and $\gamma$ are the degree of least connected node and degree exponent respectively. Hence, the average degree, $<k>$ is,
\begin{eqnarray}
  \label{kav}
   \left < k\right > &=& \int_{k_{min}}^{k_{max}}\times k \times p(k) \nonumber \\
  &=& (\gamma - 1)\times k_{min}^{\gamma -1}\times \frac{k_{max}^{2-\gamma} - k_{min}^{2-\gamma}}{2-\gamma}
\end{eqnarray}
where $k_{min}$ and $k_{max}$ are the minimum and maximum degrees exist among the nodes, respectively. For scale-free networks, the relation between minimum, and maximum degrees can be given as,
\begin{equation}
  \label{kmax}
k_{max} = k_{min}\times N^{1/(\gamma -1)}
  \end{equation}
where $N$ is the number of nodes.  Using the maximum degree relation (Eq. \ref{kmax}) in equation \ref{kav}, the average degree is given by,
\begin{eqnarray}
  \label{klarge}
  \left <  k\right > &=&  \frac{\gamma - 1}{2-\gamma}  \times \left( N^{-\frac{\gamma-2}{\gamma-1}}-1\right)\times  k_{min} \nonumber \\
  lim_{N\rightarrow \infty} \left < k \right >    &\sim& \frac{\gamma - 1 }{\gamma-2} \times  k_{min}
\end{eqnarray}
where $2 < \gamma < 3$.

$\;$

For large scale-free networks the average degree is proportional with the minimum degree, $k_{min}$. Hence, for the degree-based approach, the total number of random walks, $T\!N\!W\!$, is given by the relation, $N\!W\!P\!D \times N \times <k> $, where $N\!W\!P\!D$ is the number of random walks per degree. Even though the total number of random walks grows with the network size, $N$, the minimum degree proportionality (Eq.~\ref{klarge}) ensures the quality of embedding with minimum possible computational effort. In this approach, each edge becomes starting edge for random walks with equal probability which eliminates oversampling of the fixed number of random walk approach.

$\;$

For a fixed number of random walk cases, the number of random walks at each node must increase with the network size since the maximum degree increases with the size (Eq.~\ref{kmax}). Unless this is done, hub nodes can not be sampled accurately. To achieve the same accuracy, number of walks must increase as the size increases. Hence, the computational effort does not increase linearly with size. Moreover, starting an equal number of random walks from each node may result in creating very similar sequences for low degree nodes.  This oversampling is redundant information and becomes the source of imbalanced data at the embedding stage.

$\;$

In the natural language case, the natural flow of the sentences determines the word connectivity structure. While some words occur frequently, less common words rarely take place in the sentences.  Hence an embedding algorithm, that processes meaningful sentences, naturally distinguishes common words (central nodes, nodes with high degree) from the less common words (nodes with low degree). In the graph representation algorithms, such a natural distinction can only be imposed by the connectivity structure of the graph. Simplest and the most apparent choice is the degree of the nodes. If the number of walks which start from any node can be proportional with its degree, the appearance of these nodes will be proportional to the connectivity structure of the node. This approach has three advantages:

\begin{enumerate}
\item The nodes with a high degree will have more emphasis on the sentence-like structures, which create weighted embedded vectors. Such a degree-based approach combined with the stochastic nature of the random walk corresponds to importance sampling that improves the efficiency.
\item The abundance of low degree nodes creates an imbalanced data problem. The high contribution of low degree nodes shadows some features of the graph; by choosing degree proportionality in the number of random walks, the data imbalance problem will be eliminated.
\item Choosing degree-based random walks grossly reduces the computation time. So it can be possible to work with large-size graphs.
\end{enumerate}

The proposed model uses the node2vec strategy to visit the neighbors, with the difference that the number of walks that start from each node is proportional with the degree of the node instead of being a fixed number.

\section{Experimental Results and Analysis}

The effect of the fixed and degree-based number of random walks is tested by using three different networks. The first one of these networks is Zachary Karate Club Network \cite{girvan2002community}. This small and well-studied network is used as a test case and also a platform to discuss the implications of degree-based and the fixed number of random walks on a larger network. Two real-world networks, CORA \cite{mccallum2000automating} and CiteSeer \cite{giles1998citeseer} are chosen as test environments for the real-world networks. Node classification and link prediction are used as application areas. Table~\ref{Table1}  presents the details of the datasets used for the experimental analysis.

\begin{table}[h!]
\caption{Overview of Zachary Karate Club and two citation networks datasets.}
\begin{center}
    \begin{tabular}{|l|c|c|c|}
    \hline
        Dataset   &   $|N|$    &     $|E|$  & Classes\\   
      \hline
      Zachary Karate Club      &  34   &  78  &  2 \\ 

      \hline
      CORA      &  2,708   &  5,429  &  7 \\ 
      \hline
      CiteSeer  &  3,327   &  4,732  &  6 \\ 
      \hline
    \end{tabular}
    \label{Table1}
\end{center}
\end{table}

\subsection{Zachary Karate Club network}

Fixed and degree-based random walk approaches are employed for calculating the embedding vectors of the Zachary Karate Club network. Apart from some detailed studies \cite{perry2019detection}, the Zachary Karate Club network is commonly considered as a two community network. For comparison of the quality of embedding, with the least number of random walks,  two criteria are considered: i) cosine similarity among the embedding vectors and ii)  identification of the communities in the network using dimensional reduction techniques. For both cases, the number of walks is increased until comparable results are achieved. When the target is reached the computational efforts are compared. 

\begin{figure}[t!]
\begin{subfigure}{.5\textwidth}
  \centering
  \includegraphics[width = 8cm]{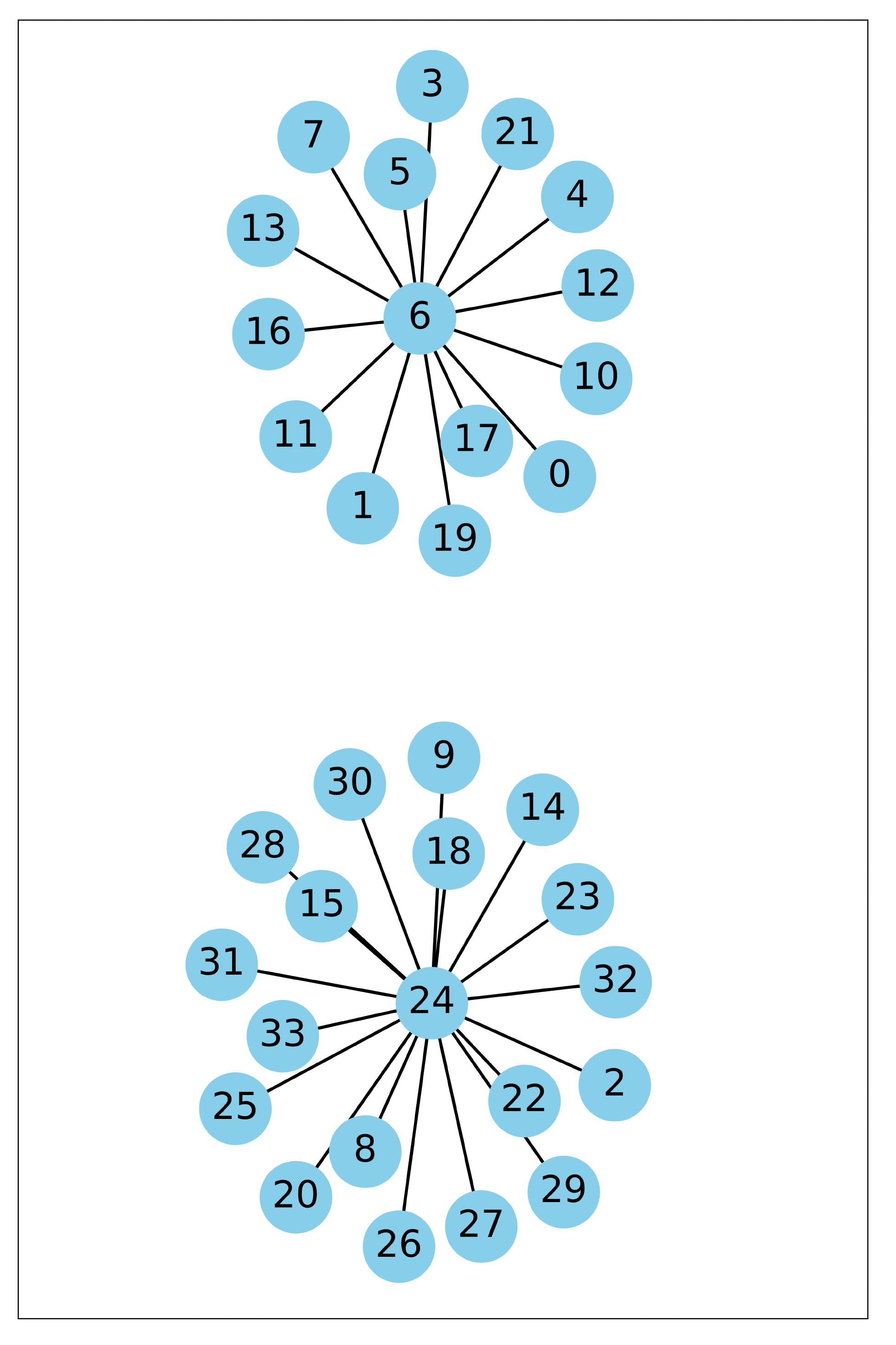}  
  \caption{Degree-based: 5 walks per degree}
  \label{cosine_N2V}
\end{subfigure}
\hfill
\begin{subfigure}{.5\textwidth}
  \centering
  \includegraphics[width = 8cm]{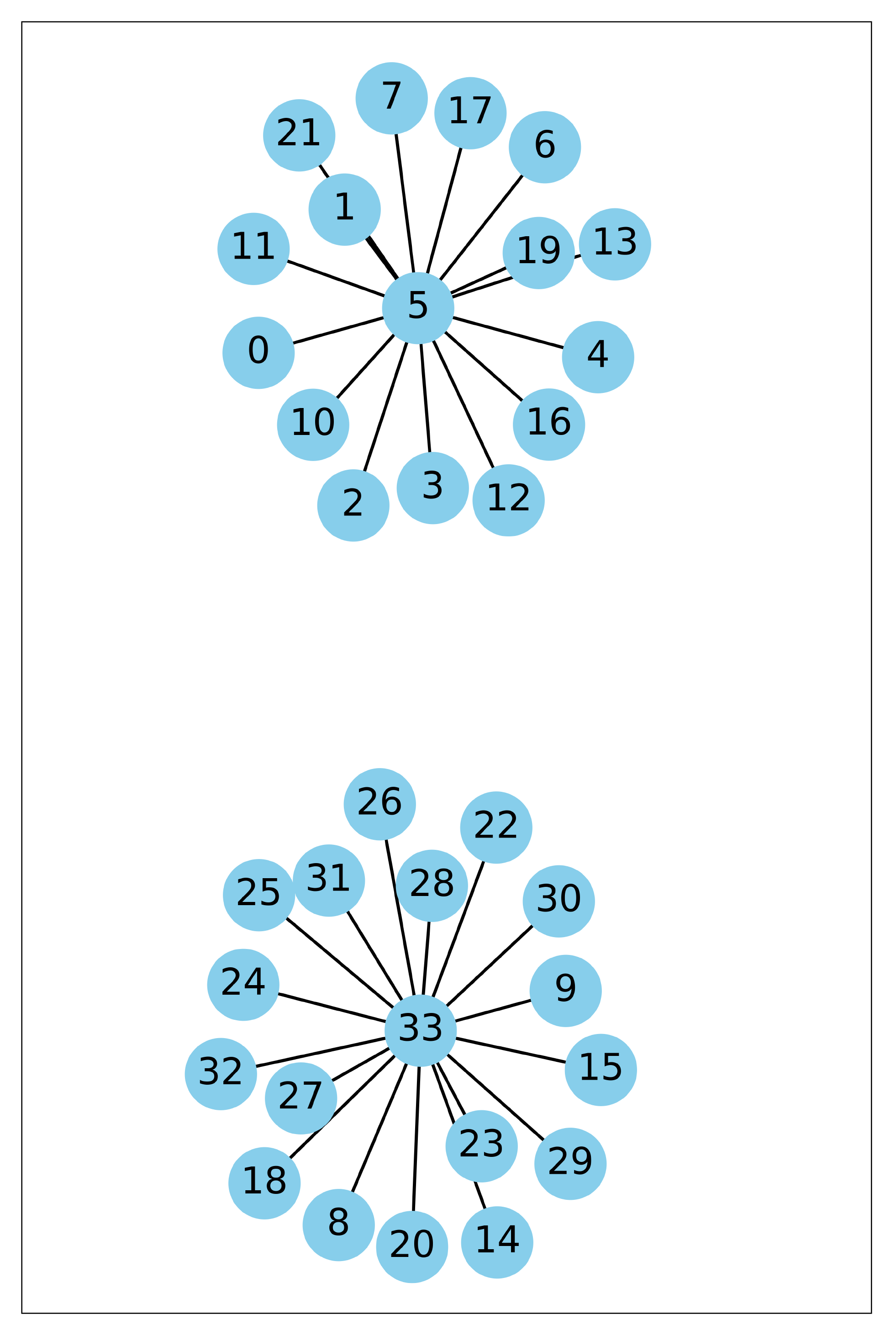}   
  \caption{Fixed-number: 40 walks per node}
  \label{cosine_DBN2V}
\end{subfigure}

\caption{Cosine similarity values (a) using $5\times {\rm degree}$ and (b) 40 random walks started at each node.}
\label{cosine}
\end{figure}

\subsubsection{Cosine similarity studies}

One way of checking the embedding quality is the identification of the
most similar node by using the cosine similarity. Embedding vectors are
calculated by keeping the window size ($WS=5$) and embedding vector
dimensionality ($32$) fixed. For each node, the cosine similarities
with the rest of the nodes are calculated regardless of being
connected or not. The nodes with maximum cosine similarity created a
topologically similar set of nodes.  Cosine similarity results are
presented by Figure \ref{cosine}. As it is seen, the
similarity calculations indicate two distinct communities. Moreover,
the communities consist of the nodes that are shown as the communities
of the Zachary Karate Club network~\cite{girvan2002community}. In the degree based case
$5$ walks per degree (Total number of $780$ walks) has been
sufficient for such separation of the communities. The same result is
achieved $40$ walks per node(Total of $1360$ walks) for the fixed
random of walks case.  The difference in the number of walks is a
clear indication of the gain in the computational time for larger
networks.

\begin{figure}[t!]
\begin{subfigure}{.5\textwidth}
  \centering
  \includegraphics[width = 8cm]{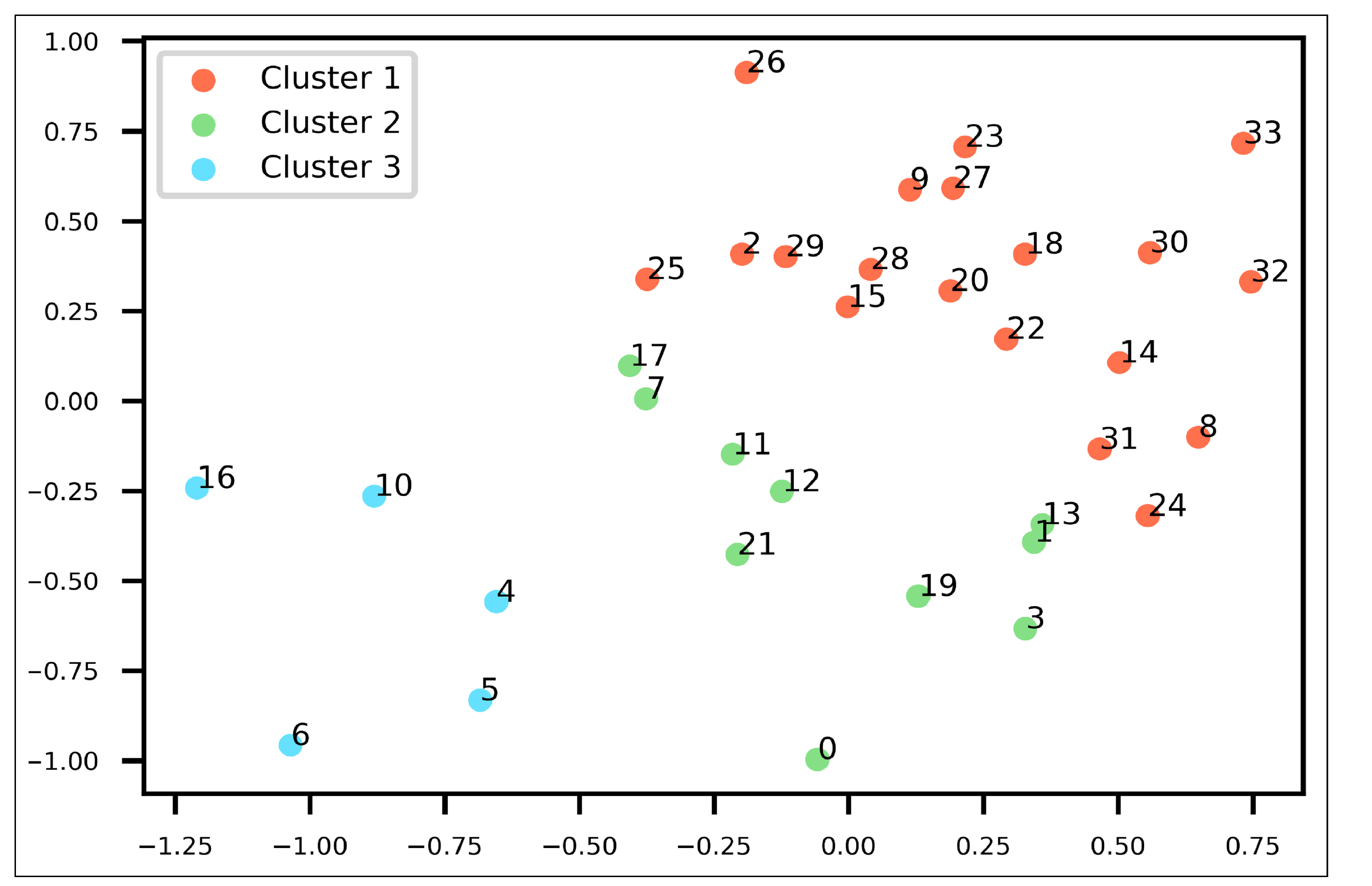}  
  \caption{Fixed number of walks}
  \label{Clustering_N2V}
\end{subfigure}
\begin{subfigure}{.5\textwidth}
  \centering
  \includegraphics[width = 8cm]{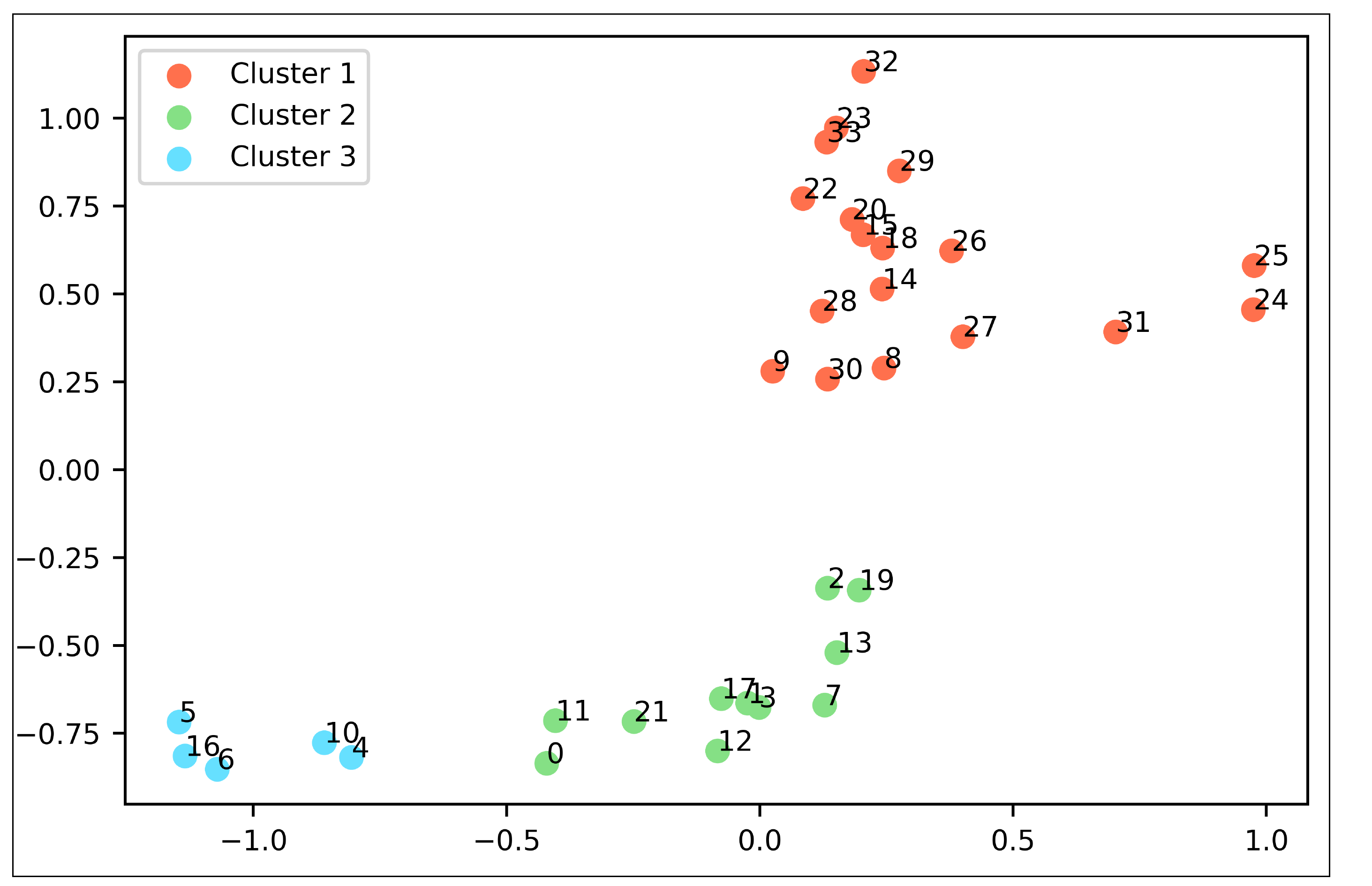}   
  \caption{Degree based walks}
  \label{Clustering_DBN2V}
\end{subfigure}

\caption{Graph representation of the Zachary Karate Club  in two dimensional representations of the nodes. The distance and colors in figure reflect the clustering. Here (a) Fixed 40   and (b)  Degree-based $5\times {\rm degree}$ number of random walks started at each node.}
\label{Clustering}
\end{figure}

\subsubsection{Community detection by using dimensional reduction }
\label{ComunityDetection}

Figure~\ref{cosine} shows that even a simple similarity study reveals the communities in accord with the well-established community structure of the network. Multidimensional Scaling (MDS) and K-Means algorithms of the python NetworkX \footnote{NetworkX developers (2021). NetworkX: Network Analysis in Python [online]. Website \url{https://networkx.org/} [accessed 17 Sep 2021].} package are used for dimensional reduction and clustering. Embedding vector space is mapped onto 2-dimensional vectors where the K-Means algorithm is used for community identification. The quality of the embedding is the assurance of the resemblance of high dimensional and low dimensional vector spaces. Figure~\ref{Clustering} shows communities and their constituent nodes. For a fixed and degree-based number of walk cases, embedding vector dimension, the walk length and window size are $32$, $10$, and $5$ respectively. For similar community identification, $1360$ walks are realized for the fixed number of random walks case while for the degree-based approach total of $5\times {\rm k_i} = 780 $ walks are observed to be sufficient.

Figures \ref{cosine} and \ref{Clustering} show that even for small graphs, the degree-based random walk approach is less computationally intensive. As the size increases, the power-law behavior of scale-free networks becomes an indicative factor for the increase of the low-degree nodes. Hence, the advantage of the degree-based approach becomes more emphasized for large networks.

\subsection{CORA and CiteSeer datasets}

CORA dataset consist of $2,708$ nodes and $5,429$ edges. Similarly, CiteSeer dataset consist of $3,327$ nodes and $4,732$ edges (Table~\ref{Table1}). By using above-mentioned datasets, embedding vectors are created using both degree-based and fixed numbers of random walks. The embedding vectors are used for i) node classification and ii) link prediction.  For node classifications, dimensional reduction and K-Means algorithms are employed similar to method explained in section~\ref{ComunityDetection}. LogisticRegressionCV program of the scikit-learn python package \cite{pedregosa2011scikit} is employed for  predictions. The link embedding vectors are constructed from the node embedding vectors by using Hadamard, L1, L2, and  Average algorithms \cite{grover2016node2vec}. During the test runs, it is observed that the Hadamard technique gives the best link prediction score. Hadamard operator is used in all link prediction calculations presented in this work.  In this experimental study, the main focus has been to observe the correlations between  accuracy, the number and lengths of random walks.

$\;$

Two parameters, the number, and length of the walks play a crucial role in mapping the topological structure of the network onto embedding. First, for fixed walk length  of $30$, the variation of accuracy values with changing number of total walks are presented in  the Tables \ref{Table2}  and  \ref{Table3}. Tables \ref{Table2}  and  \ref{Table3} present node classification accuracy values, relative percentage of total number of walks and accuracy gain with respect to $20$ runs per node case, obtained on CORA and CiteSeer datasets respectively. Tables \ref{Table2}  and  \ref{Table3} show the improvement of accuracy values with the number of walks. In the degree-based case, accuracy values increases starting from $1$ run per degree reaches a plateau at $3$ times per degree. The plateau value is also the value of the accuracy obtained by using a fixed number of walks ($20$) per node. 

\begin{table}[h!]
\caption{Summary of results in terms of node classification for degree-based node2vec and original node2vec. Number of new walks started each node depends on the degree of the node in the node based case, while for the original approach 20 new runs (fixed) for each node performed for comparison (First column of the table). For each new random walk walk continued until a  fixed walk length of 30 steps is reached. The  values are obtained by using  CORA dataset.}
\begin{center}
    \begin{tabular}{|c|c|c|c|c|}
      \hline
      Number of walks  &  Total number   &  Decrease in number  & Accuracy &  Accuracy\\
                       &  of walks (TNW) &  of walks [\%]       &          &  Loss / Gain \\
       \hline
       $1\times k_i$  &  10,138         &  79.6                &   80.3   &  -4.6   \\
        \hline
        $2\times k_i$  &  20,276         &  59                  &   83.0   &  -1.9     \\
         \hline
         $3\times k_i$  &  30,414         &  38.8                &   85.7   &  +0.8    \\
          \hline
          $4\times k_i$  &  40,552         &  18.4                &   85.0   &  +0.1    \\
           \hline
      Fixed  ($20\times n_i$)     &  49,700         &                      &   84.9   &          \\
       \hline
    \end{tabular}
  \label{Table2}
\end{center}
\end{table}

\begin{table}[h!]
\caption{Summary of results in terms of node classification for degree-based node2vec and original node2vec for the CiteSeer dataset. For number of initiated random walks and length of the random walks are the same as the CORA dataset.}
\begin{center}
    \begin{tabular}{|c|c|c|c|c|}
      \hline
      Number of walks  &  Total number   &  Decrease in number  & Accuracy &  Accuracy\\
                       &  of walks (TNW) &  of walks [\%]       &          &  Loss / Gain \\
       \hline
       $1\times k_i$      & 7,388    &    82.49   &   72.6     &   -3.4   \\
       \hline
       $2\times k_i$      & 14,776   &    64.98   &   75.5     &   -0.5   \\
       \hline
       $3\times k_i$      & 22,164   &    47.47   &   76.5     &   +0.5   \\
       \hline
       $4\times k_i$      & 29,552   &    29.97   &   76.9     &   +0.9   \\
       \hline
       $5\times k_i$     & 36,940   &    12.46   &   76.7     &   +0.7    \\
       \hline
       Fixed  ($20\times n_i$)& 42,200    &              & 76.0      &           \\
       \hline
    \end{tabular}
  \label{Table3}
\end{center}
\end{table}

$\;$

When Tables \ref{Table2} and \ref{Table3} are examined, it can be seen that the number of walks per degree is continued to increase even after accuracy is reached to a comparable value with the original algorithm. For example, for the CORA dataset Table \ref{Table2}, the original algorithm (with a fixed number (20) of walks per node) reaches $84.9 \%$ at accuracy. The Degree-based approach reaches the same accuracy between $2$- and $3$- walk per degree per node. This corresponds to a $59 \%$ to $38.8 \%$ decrease in the number of walks. Similarly, For the CiteSeer dataset Table \ref{Table3}, the original approach with fixed 20 runs per node reaches $76 \%$ accuracy. The degree-based approach has the same accuracy between $2$- and $3$- walk per degree per node. This approximately corresponds to a $64.98 \%$ to $47.74 \%$ gain in the computational time. In order to show that accuracies level off around the obtained values, the number of random walks has been increased, which reduces the gain in the computational efficiency. The computational gains and accuracies are dependent on the structure of the network and which network parameters are measured by using the obtained random walks. In the light of the above discussions, the computational gain can be predicted as an average of $50 \%$ for both CORA and CiteSeer datasets.

\begin{figure}[t!]
\begin{subfigure}{.5\textwidth}
  \centering
  \includegraphics[width = 8cm]{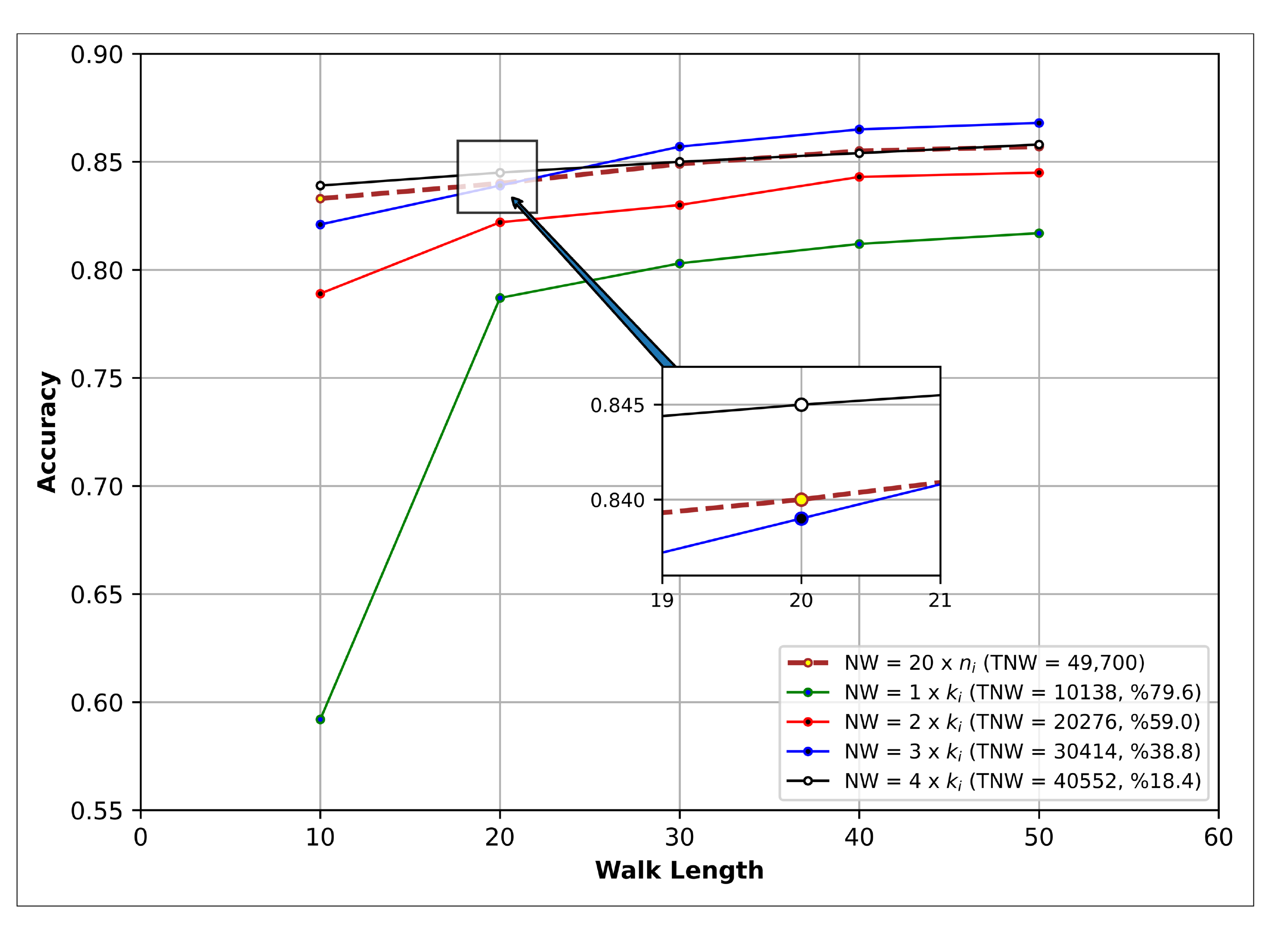}  
  \caption{Node Classification}
  \label{CORA_NodeClassification}
\end{subfigure}
\begin{subfigure}{.5\textwidth}
  \centering
  \includegraphics[width = 8cm]{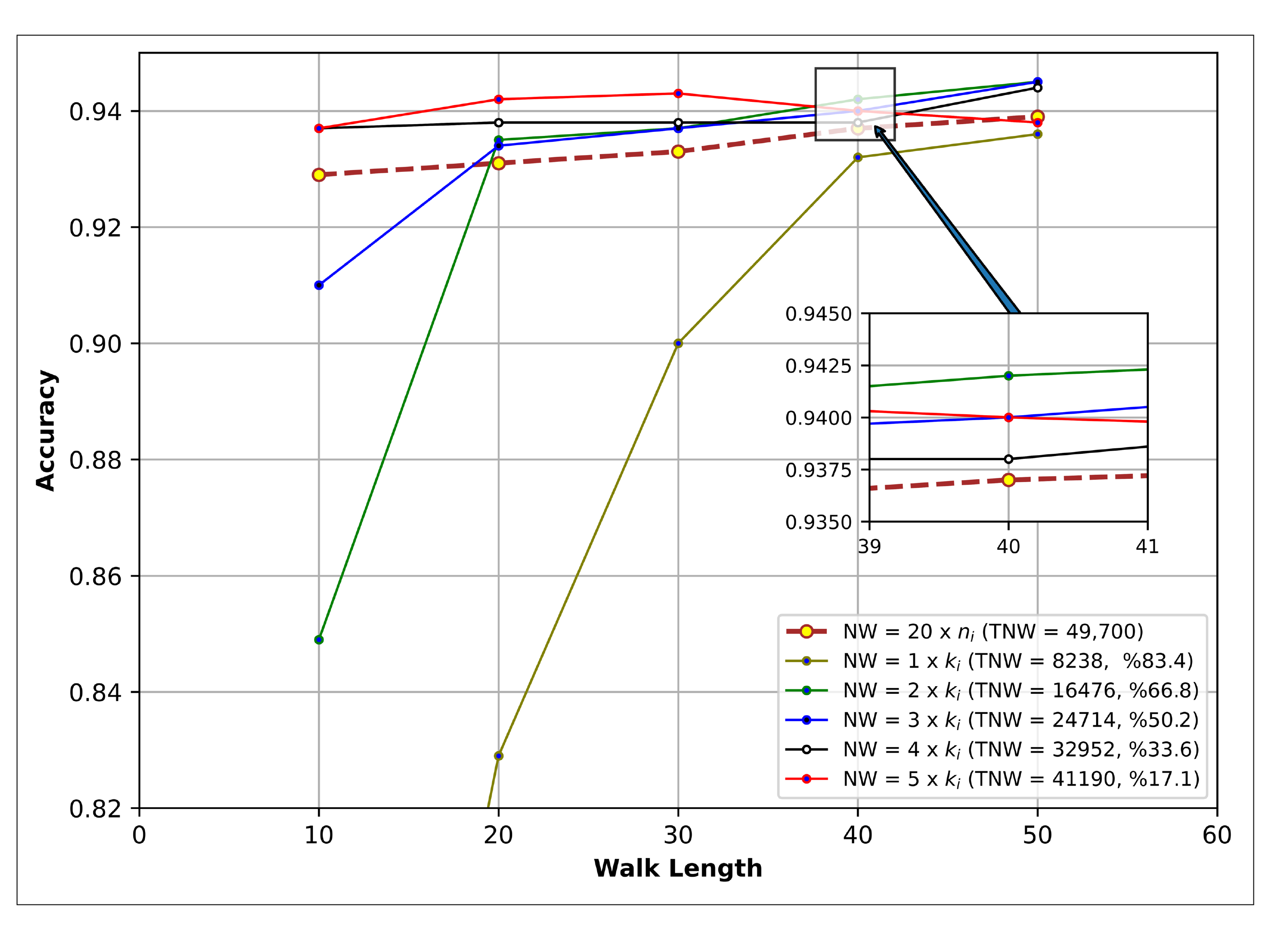}   
  \caption{Link Prediction}
  \label{CORA_LinkPrediction}
\end{subfigure}

\caption{The effects of the number and length of random walks on the accuracy of node classification and link prediction in CORA dataset.}
\label{CORA_NC_LP}
\end{figure}

$\;$

Figure~\ref{CORA_NC_LP} shows the relation between the number of walks, walk length, and accuracy for node classification (Figure~\ref{CORA_NodeClassification}), and link prediction (Figure~\ref{CORA_LinkPrediction}) for the CORA dataset. CORA network is used as a testing ground for accuracy versus walk length (10 to 50 steps). Each sub-figure (Figure~\ref{CORA_NC_LP}, a and b) contains lines representing an increasing number of random walks per node and a fixed number (20) of walks per node. From bottom to top lines indicate the increasing number of walks per degree. At a certain point accuracies of the degree-based approach and the fixed number of walks become equal. The observed general trend for a given number of walks per degree is that as the length of the walk increases, the accuracy increases. For relatively long runs a plateau can be seen. As the number of walks increases, even for short random walks results in accuracy values near the plateau values.
For the CiteSeer dataset, the observed behaviors are very similar.

$\;$

In creating a sentence-like structure using random walks, two parameters play a crucial role: Number of walks and walk length (number of steps for each new walk). For the method's efficiency, both parameters must be optimized to reach the highest accuracy with a minimal computational expense. Figure \ref{CORA_NC_LP} is devoted to such a comparative discussion. Figure \ref{CORA_NC_LP} compares degree-based and original node2vec algorithms on node classification (Figure \ref{CORA_NodeClassification}) and link prediction (Figure \ref{CORA_LinkPrediction}).

Although the identical sentence-like node sequences obtained by using the nodes of the CORA dataset are used, node classification and link prediction require different processing techniques. In the node classification case, the degree-based approach reaches and overtakes the accuracy of the original algorithm by using shorter random walks. For the CORA dataset, around 20 steps, with a number of walks higher than 2 per degree, is sufficient to reach and overpass the accuracy obtained by the original algorithm. This point can be considered as the optimized number of walk-number of steps point. For link prediction (Figure \ref{CORA_LinkPrediction}), the same situation is reached by using $40$ step long random walks. To indicate these facts, sub-figures were added into (Figure \ref{CORA_NodeClassification}) and (Figure \ref{CORA_LinkPrediction}). 

$\;$

\begin{table}[ht]
\caption{CiteSeer dataset accuracy values in terms of node classification for degree-based node2vec and original node2vec.}
\begin{center}
    \begin{tabular}{|c|c|c|c|c|c|}
\hline
                & \multicolumn{5}{c|}{Walk Length}\\ 
\hline
Number of Walks & 10 & 20 & 30 & 40 & 50\\
\hline
$1\times k_i$   & 52.3 & 67.0 & 72.6 & 73.3 &73.7 \\
\hline
$2\times k_i$   & 69.9 & 74.5 & 75.5 & 75.5 & 76.8\\
\hline
$3\times k_i$   & 72.8   &76.0  &76.5  &76.9  &77.1 \\
\hline
$4\times k_i$   & 73.2  &76.3   &76.9  &77.5  &78.6 \\
\hline
$5\times k_i$   & 74.6  &76.2   &76.7  &76.7  &77.2 \\
\hline
Fixed ($20\times n_i$)  & 75.1  & 75.8  & 76.0  & 76.7 & 77.2\\
\hline
    \end{tabular}
  \label{Table4}
\end{center}
\end{table}

\begin{table}[ht]
\caption{CiteSeer dataset accuracy values in terms of link prediction for degree-based node2vec and original node2vec.}
\begin{center}
     \begin{tabular}{|c|c|c|c|c|c|}
\hline      
& \multicolumn{5}{c|}{Walk Length}\\ 
\hline
Number of walks & 10 & 20 & 30 & 40 & 50 \\
\hline
$1\times k_i$ & 76.1 & 81.7 & 88.3 & 92.0 & 94.0\\
\hline
$2\times k_i$ & 84.9 & 93.6 & 95.1 & 95.1 & 95.5 \\
\hline
$3\times k_i$ & 93.5 & 94.9 & 95.6 & 96.4 & 96.6 \\
\hline
$4\times k_i$ & 95.4 & 96.4 & 96.3 & 96.5 & 96.3\\
\hline
$5\times k_i$ & 96.2 & 96.4 & 96.5 & 96.2 & 96.1\\
\hline
$6\times k_i$ & 96.2 & 96.2 & 96.2 & 96.2 & 95.9 \\
\hline
Fixed ($20\times n_i$) & 94.7 & 95.9 & 96.0 & 96.0 & 96.2\\
\hline
    \end{tabular}
 
  \label{Table5}
\end{center}
\end{table}

Tables~\ref{Table4} and \ref{Table5} present node classification and link prediction data for CiteSeer dataset. Starting from 20 steps of walk length, the degree-based approach reaches near plateau value accuracies as early as 3 to 4 walks per degree. The ratio between the total number of walks done for degree-based and fixed approaches indicates the efficiency of the degree-based random walk approach. Using the values presented in Tables~\ref{Table4} and \ref{Table5}, comparisons between the total number of walks done for a degree-based and fixed number of walks per node indicate that approximately $50\%$ less effort is sufficient to obtain similar accuracies.

\section{Conclusions}

Graphs provide an effective abstraction for the local and global neighborhood information on the pairwise relationships. The main drawback of graph representations is that the very valuable information can not be directly embedded into numerical vectors that can be used for further processing. There exist a large literature on the embedding methods, with different efficiencies. Matrix methods and random walk provide a basis for the majority of the proposed solutions. Among the random-walk-based solutions two of the best embedding methods, DeepWalk and node2vec make use of shallow neural network techniques which are imported from natural language processing. The sentence-like structures are created by using random walk techniques which is a computationally intensive part of the embedding process. In both of these methods, a fixed number of random walks are started from each node. The number of walks is determined by optimizing the value of the control parameter.

$\;$

In the proposed method, the number of walks is chosen proportional with the degree of the node,  instead of a fixed number for each node.   The degree-based number walk choice also has three advantages: i) High degree nodes will have more emphasis on the sentence-like structures, which will correspond to importance sampling, ii)  by choosing degree proportionality in the number of random walks, the contribution of low degree-nodes will be suppressed which eliminates the data imbalance problem, iii) degree-based random walks grossly reduces the computation time. Hence, the proposed approach is most efficient on scale-free networks.

$\;$

For large scale-free networks, the degree-based sampling approach will be more efficient since the total number of walks is proportional to the average degree (Eq.~\ref{kav}) which is a multiple of the minimum degree of the network (Eq.~\ref{klarge}).  The method realizes correct sampling of the neighborhoods while keeping the total number of walks as small as possible. Considering the majority of real-wold networks are large and scale-free, the advantage of choosing the number of walks proportional to the degree of the node becomes more apparent.  For reasonably small size lattices the proposed method is tested. Three networks, Zachary Karate Club, CORA, and CiteSeer data sets are used for node classification and link prediction. It is observed that for CORA and CiteSeer data sets, approximately $50 \%$ less computational effort than the fixed number of walks case is sufficient to reach the same accuracy for node classification and link prediction. 

Many node classification and link prediction algorithms that perform successfully are based on random walks in the literature. The idea of anonymous random walks is also based on random walks. The main future direction of our work will be to carry the idea of degree-based random walks to anonymous random walks.

\end{document}